\pdfoutput=1
\documentclass[conference]{IEEEtran}
\IEEEoverridecommandlockouts
\usepackage{fancyhdr}

\usepackage{cite}
\usepackage{amsmath,amssymb,amsfonts}
\usepackage{algorithmic}
\usepackage{graphicx}
\usepackage{textcomp}
\usepackage{url}
\usepackage{makecell}
\usepackage{array}
\usepackage{booktabs}
\usepackage{siunitx}
\usepackage{pdfpages}

\DeclareSIUnit{\transfer}{T}
\DeclareSIUnit{\op}{op}
\DeclareSIUnit{\speedup}{\ensuremath{\times}}
\newcommand{\spd}[1]{\qty[number-unit-product={}]{#1}{\speedup}}
\newcommand{\vecunits}[2]{\num{#1}$\times$\qty{#2}{\bit}}

\usepackage[table,xcdraw]{xcolor}
\def\BibTeX{{\rm B\kern-.05em{\sc i\kern-.025em b}\kern-.08em
    T\kern-.1667em\lower.7ex\hbox{E}\kern-.125emX}}

\usepackage{xcolor}
\graphicspath{{figures/}}
    
\begin{document}

\title{NVIDIA Vera: A Microarchitecture-Level Comparison with Grace, Turin, and Granite Rapids}

\author{\IEEEauthorblockN{Joseph Moore}
\IEEEauthorblockA{\textit{University of Bristol}\\
Bristol, United Kingdom \\
joseph.moore@bristol.ac.uk}
\and
\IEEEauthorblockN{Thomas Green}
\IEEEauthorblockA{\textit{University of Bristol}\\
Bristol, United Kingdom \\
thomas.green@bristol.ac.uk}
\and
\IEEEauthorblockN{Simon McIntosh-Smith}
\IEEEauthorblockA{\textit{University of Bristol}\\
Bristol, United Kingdom \\
s.mcintosh-smith@bristol.ac.uk}}

\maketitle

\thispagestyle{fancy}
\lhead{}
\rhead{}
\chead{}

\lfoot{\parbox[b]{\textwidth}{\scriptsize
\copyright~2026 IEEE. Personal use of this material is permitted. Permission from IEEE must be obtained for all other uses, in any current or future media, including reprinting/republishing this material for advertising or promotional purposes, creating new collective works, for resale or redistribution to servers or lists, or reuse of any copyrighted component of this work in other works.\\
Accepted for publication in the SC26 Workshops (PMBS26), November 15--20, 2026, Chicago, Illinois, USA.}}

\rfoot{}
\cfoot{}
\renewcommand{\headrulewidth}{0pt}
\renewcommand{\footrulewidth}{0pt}

\begin{abstract}
The majority of new HPC systems are centered around NVIDIA, AMD, or Intel CPUs. Vera is NVIDIA's latest data-center CPU, including its first custom data-center core design, Olympus. This paper investigates Vera in comparison to AMD Turin and Intel Granite Rapids, using NVIDIA Grace as a baseline. We navigate up the performance ladder, from microbenchmarks that isolate performance metrics, through NAS kernels and miniapps, to single-socket performance of a range of scientific codes.

We find that Vera achieves the best performance of the architectures evaluated in most codes tested, and that this gain is primarily attributed to its high memory bandwidth, effective Spatial Multithreading implementation, and low-latency memory subsystem aided by the architecture's predictive structures. While Vera represents a major leap over its predecessor, we show that its comparatively modest floating-point performance trails Intel and AMD in a small number of compute-bound workloads reflecting NVIDIA's GPU focus for such workloads.

\end{abstract}


\section{Introduction and related work}
CPUs compatible with the Arm ISA are no longer a novelty within HPC -- as of June 2026, four of the top ten supercomputers support the Arm ISA, two of which have the NVIDIA Grace-Hopper superchip \cite{top500}. Instruction set architecture does not determine performance in isolation; careful design of the core architecture, memory subsystem, and die topology is required to advance performance of scientific applications on CPUs. Isolating the contribution of each requires a multi-level evaluation of microbenchmarks, synthetic kernels, and real-world applications.

NVIDIA Vera is the latest data-center CPU released by NVIDIA following on from their first data-center CPU release, Grace. Vera has a monolithic compute die of 88 ``Olympus'' cores compatible with the Armv9.2-A ISA, the first data-center CPU core designed by NVIDIA.

This work analyzes the performance of a single socket of the next-generation Vera platform in comparison to current-generation CPU architectures: NVIDIA Grace, AMD Turin, and Intel Granite Rapids. We find that Vera achieves a geometric mean of \qty{1.89}{\speedup} improved performance compared to Grace across a range of HPC workloads, and outperforms the 96-core EPYC 9R45 (Turin) and Xeon 6975P-C (Granite Rapids) in the majority of codes tested. We attribute much of this speedup to advancements in the following microarchitecture structures:

\begin{itemize}
    \item The highest sustained memory bandwidth of the platforms evaluated at \qty{1088.2}{\giga\byte\per\second}.
    \item Lowest average latencies in both cache access and core-to-core communication.
    \item Spatial Multithreading that accelerates select workloads by maximizing core utilization.
    \item Prefetchers with the ability to target different patterns of irregular memory access.
\end{itemize}

CPU architecture comparison studies are often motivated by integration into a large HPC system \cite{grace-isambard}, \cite{grace-nostrum}, \cite{isambard1}, which primarily target scientific workloads or multi-node scaling performance as their main figure of merit. We take inspiration from these but extend the depth of synthetic benchmarking used, tying observed microarchitectural advancements to the performance of scientific workloads. Other works instead analyze a CPU's single-node performance more narrowly, either its microarchitecture with performance models \cite{jan-paper}, its power and energy efficiency \cite{sapphire-spec}, or a single performance factor of a specific application \cite{cloverleaf-perf-analysis}. We instead gather a holistic view of Vera, comparing performance against Grace, Turin, and Granite Rapids on a range of scientific codes, microbenchmarks, the NAS Parallel Benchmarks, and miniapps, connecting our results to understand why Vera performs as it does.

The primary contributions of our work are as follows:
\begin{itemize}
    \item The first independent microarchitectural characterization of Vera for HPC workloads to the best of our knowledge.
    \item A methodology for isolating the performance of microarchitectural features to support a holistic comparison of CPUs from different vendors.
    \item Validation of synthetic performance benchmarks against a representative suite of real-world scientific applications.
\end{itemize}

The remainder of this paper is structured as follows: We introduce the Olympus core microarchitecture in Section~\ref{overview} and our system setup in Section~\ref{setup}; Section~\ref{microbenchmarks} evaluates results of microbenchmarks that highlight the influence of the microarchitectural design; Section~\ref{synthetics} analyzes performance and scaling behavior in suites of synthetic kernels or miniapps that serve as a proxy to a wide range of HPC codes. In Section~\ref{apps} we compare the achieved performance on a representative set of real-world scientific applications. Finally, we close the paper in Section~\ref{conclusion} with our conclusions and direction for future work.

\section{Architecture Overview} \label{overview}

\begin{figure}[h]
\centerline{\includegraphics[width=1\linewidth]{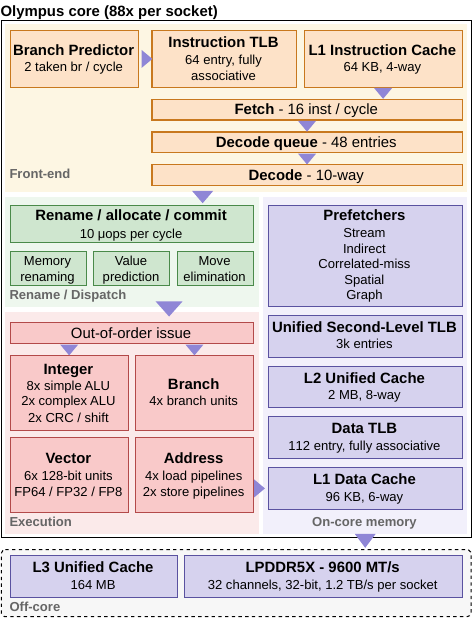}}
\caption{Abstracted block diagram of the Olympus core microarchitecture, including additional information of the off-core memory subsystem.}
\label{block-diagram}
\end{figure}

Vera is a CPU with \num{88} custom Olympus cores on a monolithic compute die. Figure~\ref{block-diagram} provides an abstract architecture block diagram of the core pipeline.

Olympus provides a wide front end relative to other architectures, with a fetch width of \num{16} instructions per cycle and \num{10}-way decode which can include fused instructions, backed by a branch predictor capable of predicting two taken branches per cycle. Grace's Neoverse-V2 cores sustain six micro-ops per cycle \cite{neoverse-v2-chips}, Intel's Redwood Cove can fetch \num{32} bytes per cycle and also sustains six micro-ops per cycle \cite{redwood-chips}, while AMD's Zen 5 can provide \num{64} bytes per cycle aggregated bandwidth via two ports and can sustain eight micro-ops per cycle \cite{zen5-hotchips}.

Olympus supports Arm's Scalable Vector Extension (SVE) and NEON vector instructions and has six 128-bit SVE units. Each core is equipped with a \qty{96}{\kibi\byte} L1d, \qty{64}{\kibi\byte} L1i, and \qty{2}{\mebi\byte} unified L2 cache. All cores share a single \qty{164}{\mebi\byte} unified L3 cache. Vera's main memory is provided by LPDDR5X memory modules. NVIDIA discloses five classes of prefetcher on Olympus: Stream, Indirect, Correlated-Miss, Spatial, and Graph.

Vera improves on the high main-memory bandwidth of Grace, extending Grace's 32 16-bit memory channels running at a measured \qty{8532}{\mega\transfer\per\second} to Vera's 32 32-bit channels running at \qty{9600}{\mega\transfer\per\second}. This leads to a peak main-memory bandwidth of $9600\ \text{MT/s} \times \frac{32\ \text{bit}}{8\ \text{bit/byte}} \times 32\ \text{channels} = 1228.8\ \text{GB/s}$. This represents a \qty{2.25}{\speedup} increase over the Grace-Grace superchip's theoretical per-socket peak bandwidth of \qty{546.0}{\giga\byte\per\second}.

Vera implements Simultaneous Multithreading (SMT) via Spatial Multithreading with two threads per core for a total of 176 threads. Spatial Multithreading partitions a core's pipeline resources between threads physically. Intel's Hyper-Threading \cite{intel-hyperthread} duplicates the architectural state and splits some buffers, such as the reorder buffer and the load/store buffers, between the two threads, but the threads share the execution units dynamically. Instead, Olympus splits nearly all pipeline resources between both threads, meaning that they mainly share the caches. This removes contention between threads for execute units, but also limits the pipeline width of each thread to half of the core's width when both threads are active. Grace has no SMT implementation, while Turin does but it is not exposed as an option on the node tested in this work, hence limiting the generalization of our SMT analysis for Turin. For the remainder of this paper, ``SMT on'' is used to indicate that two software threads are assigned per physical core, while ``SMT off'' is used to indicate only one thread per core.

Olympus dynamically detects whether the core is in single-threaded or multi-threaded mode depending on whether the second thread is idle. In single-threaded mode, the single thread has access to all core resources. Upon a switch, nearly all pipeline resources are halved. In order to avoid idle vector units where unnecessary, the option to have vector execution units only split when both threads are executing either SVE or NEON instructions is available in the BIOS. The branch prediction unit also splits between threads, with a dual ported unit easily separable into a single port per thread.

\section{Experimental Setup} \label{setup}

This paper benchmarks a single socket of the Vera CPU, comparing against the previous generation NVIDIA Grace CPU, as well as an AMD 5th-Generation ``Turin'' CPU and an Intel 6th-generation Xeon ``Granite Rapids'' CPU. We test an early access single-socket engineering sample of Vera that represents the Vera CPU dual-socket server product, so dual-socket or CPU-GPU performance analysis was not available at this time. The air-cooled engineering board was artificially power capped to achieve a \qty{450}{\watt} TDP equivalent.

We compare against NVIDIA Grace as part of the Isambard 3 supercomputer at the Bristol Centre for Supercomputing. Each of the 384 nodes is equipped with the NVIDIA Grace Superchip containing two Grace CPUs per node, connected with the NVIDIA NVLink-C2C interconnect. Each CPU has access to \qty{120}{\giga\byte} of local LPDDR5X memory for a total of \qty{240}{\giga\byte} of memory per node. As Isambard 3 has dual-socket Grace CPUs we take care throughout our benchmarking to pin all cores and memory allocations to a single socket.

Our comparisons to Turin and Granite Rapids CPUs are performed on compute-optimized bare-metal AWS instances for both architectures. Both instances have cloud-specific SKUs with 96 physical cores per socket each. While both the Turin \mbox{c8a.metal-24xl} and Granite Rapids \mbox{c8i.metal-48xl} are single-socket instances, the Intel instance has additional vCPUs due to two Hyper-Threads per core, while the Turin SKU has no multithreading. We once again take care to only assign one thread per physical core when we are not testing multi-threaded behavior. These SKUs were chosen due to availability of hardware and similar core counts to Vera (88 versus 96). SKUs with greater core counts exist in both families, with more memory channels available in different Granite Rapids CPUs. Our results therefore only compare performance to the 9R45 and 6975P-C.

\begin{table}[t]
\centering
\caption{System configurations of CPUs evaluated.}
\label{tab/sysconfig}
\renewcommand{\arraystretch}{1.15}
\setlength{\tabcolsep}{4pt}
\footnotesize
\begin{tabular}{@{}>{\raggedright\arraybackslash}p{2.0cm}
                  >{\centering\arraybackslash}p{0.9cm}
                  >{\raggedright\arraybackslash}p{1.8cm}
                  >{\raggedright\arraybackslash}p{2.7cm}@{}}
\toprule
\rowcolor[HTML]{D9D9D9}
\textbf{CPU} & \textbf{Cores / Threads} & \textbf{System} & \textbf{OS / Kernel} \\
\midrule
NVIDIA Vera & 88 / 176 & Engineering Sample & Ubuntu 24.04.3, 6.17.0-1017-\allowbreak nvidia-64k \\
\rowcolor[HTML]{F2F2F2}
NVIDIA Grace & 72 / 72 & Isambard 3 & SLES 15 SP7, 6.4.0-150700.\allowbreak 53.66-64kb \\
AMD EPYC 9R45 (Turin) & 96 / 96 & AWS c8a.\allowbreak metal-24xl & Ubuntu 24.04.4, 6.17.0-1017-aws \\
\rowcolor[HTML]{F2F2F2}
Intel Xeon 6975P-C (Granite Rapids) & 96 / 192 & AWS c8i.\allowbreak metal-48xl & Ubuntu 24.04.4, 6.17.0-1017-aws \\
\bottomrule
\end{tabular}
\end{table}

Table~\ref{tab/sysconfig} reports the system configurations used in testing. We use GCC 15.3.0 for all tests, and align OS and kernel versions where possible.

\section{Microbenchmarks} \label{microbenchmarks}

\subsection{Sustained Memory Bandwidth}

We measure sustained memory bandwidth using the STREAM microbenchmark \cite{stream}. STREAM measures achieved memory bandwidth using four kernels - Copy, Scale, Add, and Triad, each performing a mix of loads, stores, and arithmetic. We utilize triad as our primary figure of merit as it best reflects the achievable bandwidth in codes performing arithmetic on loaded data as with many HPC codes.

All arrays are sized according to the cache sizes of the platform STREAM is tested on, ensuring each array requires at least \num{4}$\times$ the size of the last-level cache. For Vera, we use an array size of 183,000,000 which equates to \qty{1.36}{\gibi\byte} per array.

When running across all cores, we use \texttt{OMP\_PROC\_BIND=spread} and \texttt{OMP\_PLACES=cores} with one thread per core. We do not observe that distributing thread placement leads to more performance when measuring memory bandwidth with all cores active, but we specify it for a deterministic, reproducible thread-to-core mapping. 

As we only want to compare to a single Vera socket while our Grace implementation is a dual-socket machine, on Grace we run with \texttt{numactl --cpunodebind=0 --membind=0} to pin all threads to a single socket and allocate to the memory closest to that socket. Neither Vera nor Grace have multiple NUMA domains per socket, and our Turin node is configured with a single NUMA domain per socket. Our Granite Rapids instance has three NUMA domains per socket and thus relies on first-touch allocation within STREAM's initialization step.

For scaling the number of cores, we increase the available pool of cores via \texttt{taskset} with a stride of 8 between logical CPU IDs. While distant logical cores do not guarantee distant physical cores, we find that it aids the scaling performance by spreading out cores faster. We choose 8 as this is the size of a single CCD in our Turin SKU, thus spreading cores evenly across CCDs. This also benefits Intel as it more quickly fills cores across each of its three compute chiplets. We find that compared to sequentially incrementing core ID that this does not have a performance impact on Vera or Grace, but has a significant scaling impact on Turin due to the CCD-based topology and on Intel with their coarser chiplet-based architecture. Where Turin and Granite Rapids require careful pinning across applications to maximize memory bandwidth, Vera provides this regardless of core placement.

\begin{figure}[t]
  \centering
  \input{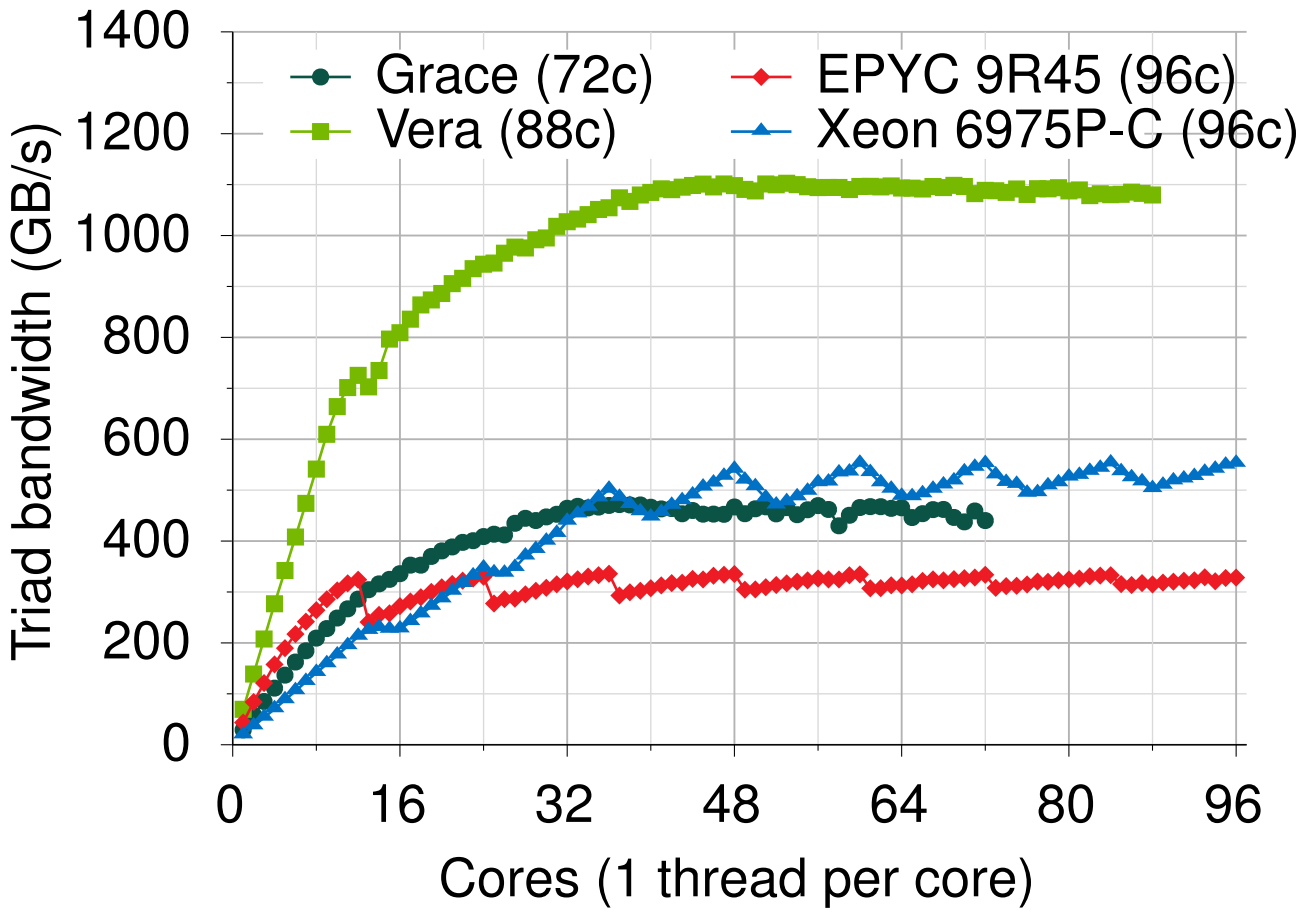}
  \caption{Sustained STREAM triad memory bandwidth of different CPUs, scaling
  across the total number of physical cores. Core placement is spread in a
  stride of 8 with a round-robin pattern.}
  \label{stream-rr}
\end{figure}

Figure~\ref{stream-rr} demonstrates that Vera achieves the highest sustained memory bandwidth at \qty{1088.2}{\giga\byte\per\second} in an all-core run, reaching \qty{88.6}{\percent} of the theoretical peak bandwidth. In comparison, Grace achieved \qty{449.4}{\giga\byte\per\second} which is \qty{82.3}{\percent} of its theoretical peak. We cannot calculate a fixed theoretical peak  memory bandwidth for our Granite Rapids and Turin CPUs as it can depend on both the number of DIMMs as well as the MT/s rate of those DIMMs. In this case, the Intel Xeon 6975P-C achieved \qty{552.6}{\giga\byte\per\second} while the AMD EPYC 9R45 achieved \qty{332.1}{\giga\byte\per\second}. We measure that Vera had \qty{1.97}{\speedup} greater main-memory bandwidth than the next best CPU which was Granite Rapids. Notably, Vera's per-core contribution of memory bandwidth at few cores is the largest out of the comparison architectures. At 12 cores, Vera achieves \qty{725.5}{\giga\byte\per\second} -- \qty{2.54}{\speedup} that of Grace, \qty{2.24}{\speedup} that of Turin, and \qty{3.41}{\speedup} that of Granite Rapids.

\subsection{Cache latencies}

We measure the latency of memory accesses that correspond to the capacity of each level of cache through the test-tlb benchmark \cite{test-tlb}. This microbenchmark performs pointer chasing across a fixed size array, allowing the timing of a chain of indirect loads within a working set sized to fit into different levels of cache. One random permutation of memory accesses is repeated for at least \qty{0.2}{\second}. This setup is repeated for five trials with new permutations each time, with the minimum of these reported. These results are further averaged over multiple trials for most array sizes, with ten repeats for the shortest arrays and scaling down to one for the largest to balance experiment runtime. The random-access pattern is designed to circumvent stride prefetchers, while utilizing huge pages reduces TLB latency that would otherwise obscure our results. This benchmark is intended to be used to measure the latency of cache accesses, but many factors including prefetching, page sizes, and huge page allocation, can change the result from the true cache latency to one specific to this memory access pattern.

We use a stride of 64 bytes, though as the array is randomly permuted, this instead sets the spacing between array elements. On Vera, we run this experiment with the recommended \qty{64}{\kibi\byte} page size in addition to \qty{2}{\mebi\byte} huge pages. We do not use the recommended \qty{512}{\mebi\byte} huge pages on Vera to make a fair comparison with Grace, where we could not modify huge page sizing. Grace was also fixed to a \qty{64}{\kibi\byte} page size and \qty{2}{\mebi\byte} huge pages enabled, while the remaining CPUs had \qty{4}{\kibi\byte} pages and \qty{2}{\mebi\byte} huge pages.

\begin{figure}[t]
  \centering
  \input{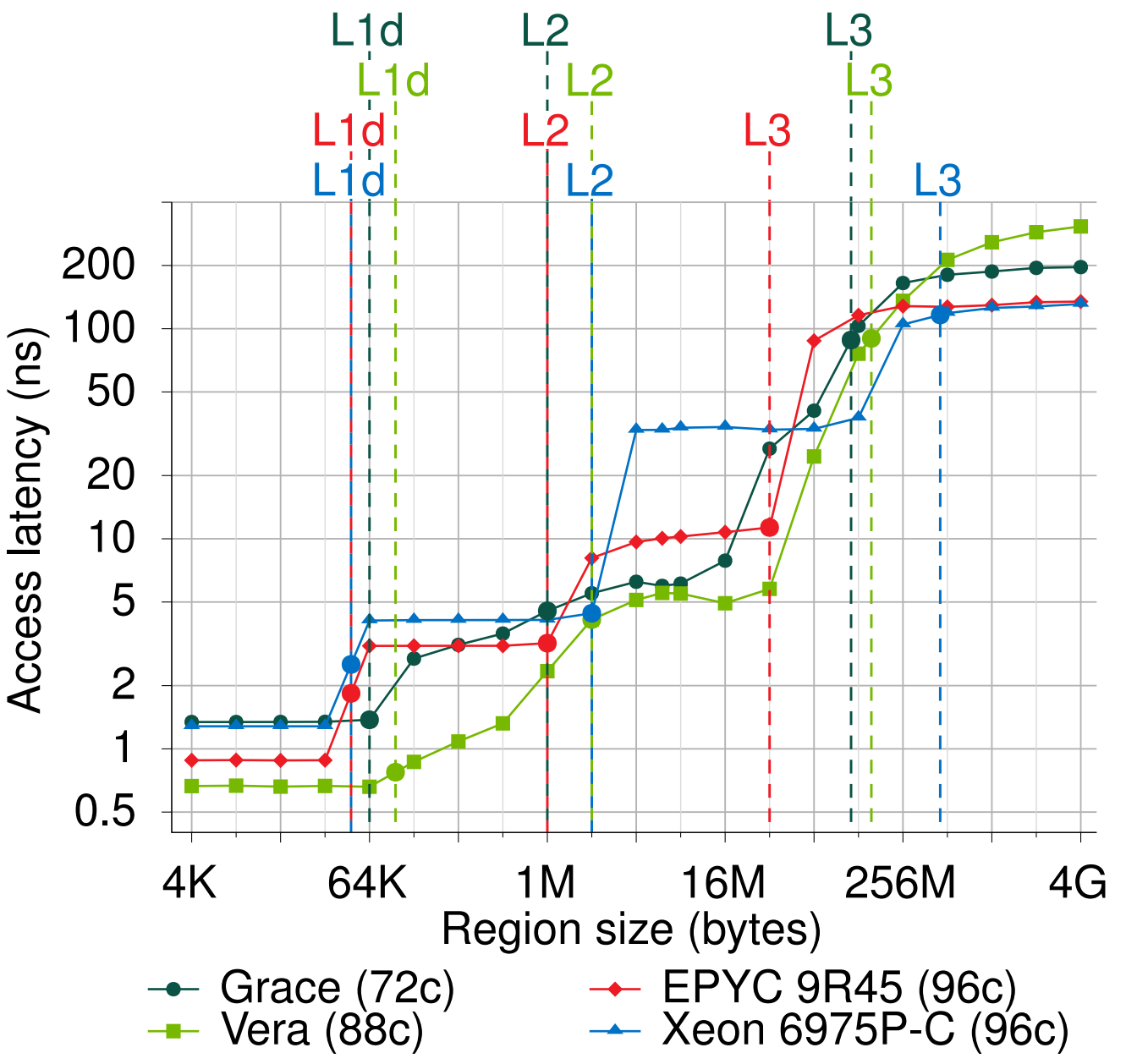}
  \caption{Measured time per dependent load (ns) across different region
  sizes, measured in a random-order pointer chasing sequence. \qty{2}{\mebi\byte} huge pages enabled to reduce additional TLB latency. Dashed lines mark the L1d, L2 and L3 sizes of each CPU; where two CPUs have the same size, the line alternates color.}
  \label{cache-latency}
\end{figure}

Figure~\ref{cache-latency} compares the measured access latencies at different region sizes for each of our CPUs. We find that Vera achieves the lowest time per load at region sizes lower than \qty{128}{\mebi\byte}, though we observe unexpected results for both Grace and Vera. The step function presented for Turin and Granite Rapids is what we would expect to see for a random-access pattern, with the latency of each access spiking once our region is larger than the previous layer of cache, before maintaining roughly the same latency. Instead, once the region size is greater than the size of L1d cache, the latency is lower than expected, increases gradually within each cache level, and does not spike sharply upon a cache boundary.

Frumusanu observed a similar pattern in \cite{anandtech-latency} for the Cortex-X1 and Cortex-A78 core designs between the L2 and L3 caches. They showed that this is only observable in repeated pointer-chasing paths and was the effect of the temporal prefetcher which is documented to exist in other Arm cores such as the Neoverse-N2 \cite{hotchips-temporal}. Grace's Neoverse-V2 cores show the same behavior as these cores, which are confirmed to have a temporal prefetcher. The correlated-miss prefetcher that is confirmed to be present in Vera is an implementation of a temporal prefetcher \cite{vera-whitepaper}. This class of prefetcher differs from a stride prefetcher, which relies on fixed deltas between memory addresses, by exploiting recurrences in access patterns by recording the memory addresses in a chain of misses, allowing them to be replayed in the event of a recurrence. A temporal prefetcher can therefore achieve lower latency on a random-access memory pattern if the same memory address path is repeated across iterations. In test-tlb, a correlated-miss prefetcher would be effective as the same permutation of memory accesses is repeated after one untimed warm-up run.

On Vera, the time per load stays close to the L1d value beyond the L1d size, then rises gradually up to \qty{2}{\mebi\byte}. The timing at the size of the L2 appears to expand beyond Grace's lower latency region of \qty{16}{\mebi\byte} to \qty{32}{\mebi\byte}. A load that misses L2 cannot return in L2 hit time, so the cache line must have been prefetched. Note that unlike Grace, the effect of the correlated-miss prefetcher is observed beyond L1d sizes instead of only beyond L2. We could not disable individual prefetchers on Vera, so Vera's true L1d and L2 hit latencies are not measured.

Figure~\ref{cache-latency} therefore shows that up to a \qty{128}{\mebi\byte} region, Vera has the lowest time per load of the four CPUs. Codes that have repeated irregular accesses, such as sparse iterative solvers or unstructured-mesh codes, may see lower effective latency compared to Turin or Granite Rapids which demonstrate no equivalent prefetching in this benchmark. While NVIDIA lists several prefetchers for different memory access patterns, if a pattern is not prefetchable or exceeds the size of the L3, this benefit is lost. Above \qty{512}{\mebi\byte}, Vera's main-memory latency is the largest out of the CPUs tested.

\subsection{Core-to-Core latency}

Core-to-core latency is measured through timing the cache-coherency protocol when one core writes to a shared cache line and another core immediately attempts to read. This is implemented in a microbenchmark written in Rust, with commit \textit{2592303} compiled on all CPUs with a rustc version newer than 1.90.0 (October 2025) \cite{c2c-latency}. Each measurement is half the mean round-trip time over 300 samples of 1000 round trips after one discarded warm-up. The test is symmetric, so each pair of core IDs is measured once. Core IDs are pinned to one hardware thread per core.

\begin{figure}[t]
  \centering
  \input{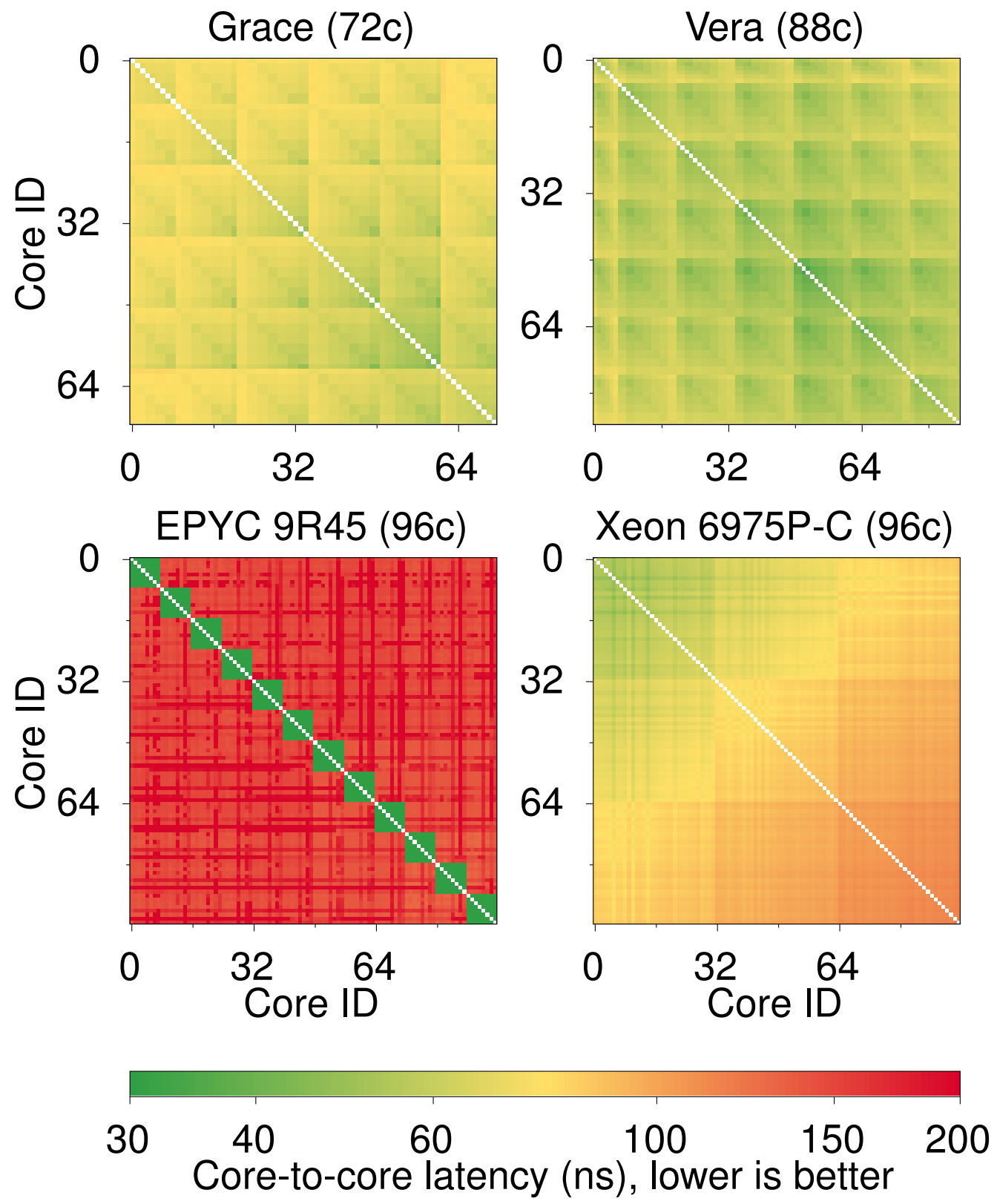}
  \caption{Core-to-core latencies (\unit{\nano\second}) of each pair of cores on different CPUs. Results measure the timing of the cache-coherency protocol for a shared cache line between two cores. A logarithmic latency color-scale is used.}
  \label{c2c-comparisons}
\end{figure}

\begin{figure}[t]
  \centering
  \input{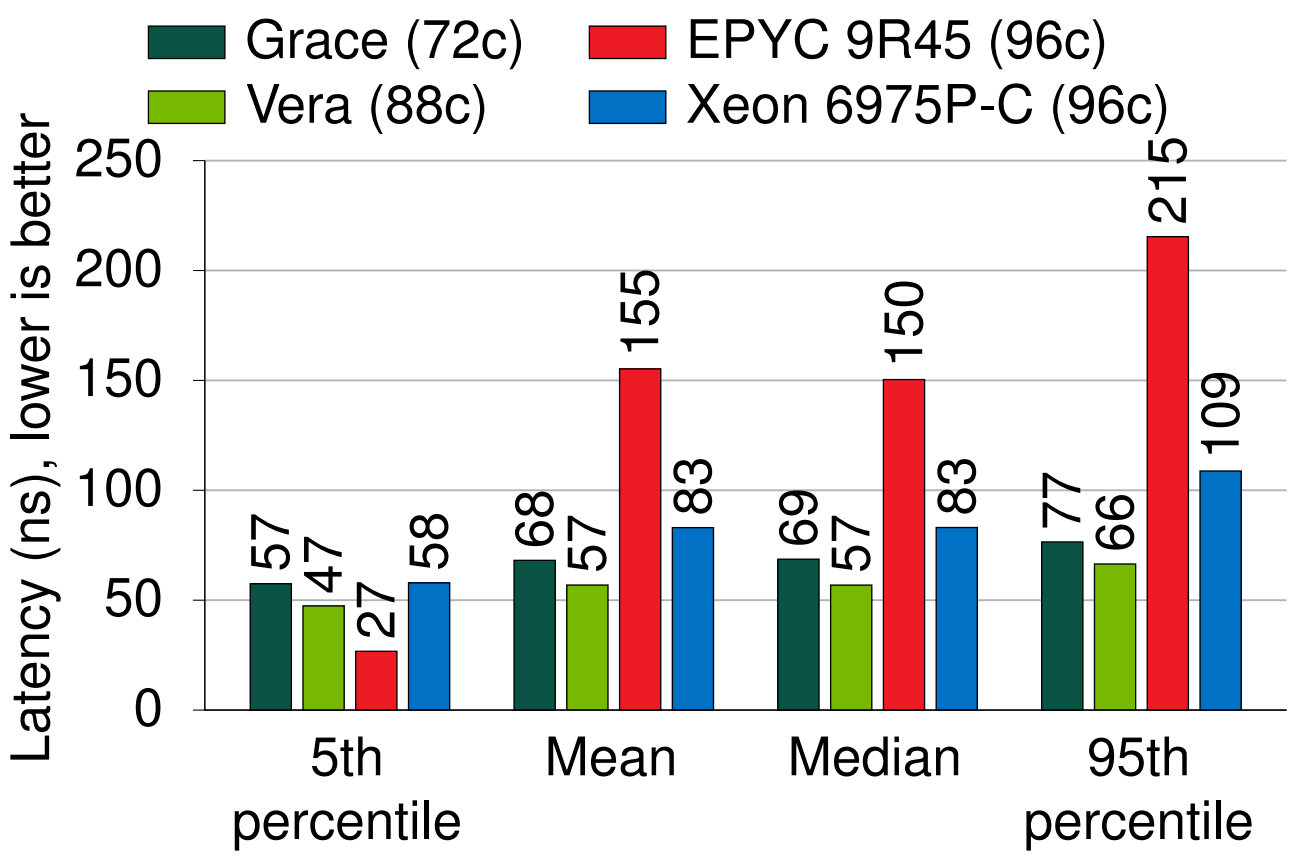}
  \caption{Distribution of core-to-core latencies (\unit{\nano\second}) across all core pairs per CPU. Results derived from the same measurements shown in Figure~\ref{c2c-comparisons}.}
  \label{c2c-stats}
\end{figure}

Figure~\ref{c2c-comparisons} presents the measured latencies for all pairs of cores on each CPU, while Figure~\ref{c2c-stats} provides further statistics on the latency distribution. While logical core IDs do not necessarily equate to fabric distance between pairs of cores, the patterns of the results often imply fabric topology. We see similar checkerboard patterns on NVIDIA Grace and Vera, implying that the latency is distance-based in a 2D mesh-like fabric. Vera has a reduction of \qty{18}{\percent} in 5th percentile, \qty{17}{\percent} in mean, \qty{17}{\percent} in median, and a \qty{13}{\percent} reduction in 95th percentile core-to-core latencies compared to Grace despite having \qty{22}{\percent} more cores, implying a larger compute die and thus greater distance between some core pairs.

Vera also has lower core-to-core latencies than Granite Rapids, with a \qty{31}{\percent} reduction in mean latencies, \qty{18}{\percent} in the 5th percentile, and \qty{39}{\percent} in the 95th percentile. Granite Rapids' three compute dies are visible in the heatmap, with latencies between \qty{45}{\nano\second} and \qty{61}{\nano\second} within a chiplet, increasing by approximately \qty{14}{\nano\second} for each hop to other chiplets. As two hops are required to the farthest chiplet, we observe an extra \qty{28}{\nano\second} of latency for a single core in the farthest chiplet, or doubled to \qty{56}{\nano\second} if both cores are resident in the farthest compute chiplet from the cache line. Within a single compute chiplet, there is a range of approximately \qty{20}{\nano\second} which is similar to that of Vera across its single die.

Turin has a unique chiplet architecture with 8-core CCDs, leading to the lowest latencies out of any of the CPUs at approximately \qty{27}{\nano\second} within a CCD. Latency is significantly higher than that of the other CPUs once communicating between cores in different CCDs, ranging between \qty{130}{\nano\second} and \qty{244}{\nano\second} with no clear pattern. In the best case, this is \spd{1.69} Vera's worst latency of \qty{77}{\nano\second}, and thus we expect that any latency-bound multi-core codes will exhibit poorer scaling on Turin when scaling beyond a single CCD compared to Vera which has more consistent and lower average latencies.

\section{Synthetic codes} \label{synthetics}

\subsection{NAS Parallel Benchmarks}
The NAS Parallel Benchmarks are a suite of kernels and pseudo-applications derived from CFD codes that present mixed performance profiles \cite{nas}. They are written to have no external dependencies, and the main release that we use contains no architecture-specific optimizations. We run all original eight benchmarks (split into five kernels and three pseudo-applications), as well as UA as a benchmark for irregular memory accesses.

\begin{figure*}[t]
  \centering
  \begingroup
\fontfamily{phv}\selectfont
  \makeatletter
  \providecommand\color[2][]{%
    \GenericError{(gnuplot) \space\space\space\@spaces}{%
      Package color not loaded in conjunction with
      terminal option `colourtext'%
    }{See the gnuplot documentation for explanation.%
    }{Either use 'blacktext' in gnuplot or load the package
      color.sty in LaTeX.}%
    \renewcommand\color[2][]{}%
  }%
  \providecommand\includegraphics[2][]{%
    \GenericError{(gnuplot) \space\space\space\@spaces}{%
      Package graphicx or graphics not loaded%
    }{See the gnuplot documentation for explanation.%
    }{The gnuplot epslatex terminal needs graphicx.sty or graphics.sty.}%
    \renewcommand\includegraphics[2][]{}%
  }%
  \providecommand\rotatebox[2]{#2}%
  \@ifundefined{ifGPcolor}{%
    \newif\ifGPcolor
    \GPcolortrue
  }{}%
  \@ifundefined{ifGPblacktext}{%
    \newif\ifGPblacktext
    \GPblacktextfalse
  }{}%
  \let\gplgaddtomacro\g@addto@macro
  \gdef\gplbacktext{}%
  \gdef\gplfronttext{}%
  \makeatother
  \ifGPblacktext
    \def\colorrgb#1{}%
    \def\colorgray#1{}%
  \else
    \ifGPcolor
      \def\colorrgb#1{\color[rgb]{#1}}%
      \def\colorgray#1{\color[gray]{#1}}%
      \expandafter\def\csname LTw\endcsname{\color{white}}%
      \expandafter\def\csname LTb\endcsname{\color{black}}%
      \expandafter\def\csname LTa\endcsname{\color{black}}%
      \expandafter\def\csname LT0\endcsname{\color[rgb]{1,0,0}}%
      \expandafter\def\csname LT1\endcsname{\color[rgb]{0,1,0}}%
      \expandafter\def\csname LT2\endcsname{\color[rgb]{0,0,1}}%
      \expandafter\def\csname LT3\endcsname{\color[rgb]{1,0,1}}%
      \expandafter\def\csname LT4\endcsname{\color[rgb]{0,1,1}}%
      \expandafter\def\csname LT5\endcsname{\color[rgb]{1,1,0}}%
      \expandafter\def\csname LT6\endcsname{\color[rgb]{0,0,0}}%
      \expandafter\def\csname LT7\endcsname{\color[rgb]{1,0.3,0}}%
      \expandafter\def\csname LT8\endcsname{\color[rgb]{0.5,0.5,0.5}}%
    \else
      \def\colorrgb#1{\color{black}}%
      \def\colorgray#1{\color[gray]{#1}}%
      \expandafter\def\csname LTw\endcsname{\color{white}}%
      \expandafter\def\csname LTb\endcsname{\color{black}}%
      \expandafter\def\csname LTa\endcsname{\color{black}}%
      \expandafter\def\csname LT0\endcsname{\color{black}}%
      \expandafter\def\csname LT1\endcsname{\color{black}}%
      \expandafter\def\csname LT2\endcsname{\color{black}}%
      \expandafter\def\csname LT3\endcsname{\color{black}}%
      \expandafter\def\csname LT4\endcsname{\color{black}}%
      \expandafter\def\csname LT5\endcsname{\color{black}}%
      \expandafter\def\csname LT6\endcsname{\color{black}}%
      \expandafter\def\csname LT7\endcsname{\color{black}}%
      \expandafter\def\csname LT8\endcsname{\color{black}}%
    \fi
  \fi
    \setlength{\unitlength}{0.0500bp}%
    \ifx\gptboxheight\undefined%
      \newlength{\gptboxheight}%
      \newlength{\gptboxwidth}%
      \newsavebox{\gptboxtext}%
    \fi%
    \setlength{\fboxrule}{0.5pt}%
    \setlength{\fboxsep}{1pt}%
    \definecolor{tbcol}{rgb}{1,1,1}%
\begin{picture}(10280.00,3740.00)%
    \gplgaddtomacro\gplbacktext{%
      \csname LTb\endcsname%
      \put(644,801){\makebox(0,0)[r]{\strut{}0.5}}%
      \csname LTb\endcsname%
      \put(644,1630){\makebox(0,0)[r]{\strut{}1}}%
      \csname LTb\endcsname%
      \put(644,2460){\makebox(0,0)[r]{\strut{}2}}%
      \csname LTb\endcsname%
      \put(644,3289){\makebox(0,0)[r]{\strut{}4}}%
      \csname LTb\endcsname%
      \put(1264,615){\makebox(0,0){\strut{}BT}}%
      \csname LTb\endcsname%
      \put(2125,615){\makebox(0,0){\strut{}CG}}%
      \csname LTb\endcsname%
      \put(2986,615){\makebox(0,0){\strut{}EP}}%
      \csname LTb\endcsname%
      \put(3847,615){\makebox(0,0){\strut{}FT}}%
      \csname LTb\endcsname%
      \put(4709,615){\makebox(0,0){\strut{}IS}}%
      \csname LTb\endcsname%
      \put(5570,615){\makebox(0,0){\strut{}LU}}%
      \csname LTb\endcsname%
      \put(6431,615){\makebox(0,0){\strut{}MG}}%
      \csname LTb\endcsname%
      \put(7292,615){\makebox(0,0){\strut{}SP}}%
      \csname LTb\endcsname%
      \put(8153,615){\makebox(0,0){\strut{}UA}}%
      \csname LTb\endcsname%
      \put(9316,615){\makebox(0,0){\strut{}\raisebox{0pt}[\ht\strutbox][\dp\strutbox]{\begin{tabular}[t]{@{}c@{}}Geo. mean\\(best of SMT off/on)\end{tabular}}}}%
    }%
    \gplgaddtomacro\gplfronttext{%
      \csname LTb\endcsname%
      \put(9140,2593){\rotatebox{90.00}{\makebox(0,0)[l]{\strut{}2.13}}}%
      \csname LTb\endcsname%
      \put(9347,1880){\rotatebox{90.00}{\makebox(0,0)[l]{\strut{}1.18}}}%
      \csname LTb\endcsname%
      \put(9553,1854){\rotatebox{90.00}{\makebox(0,0)[l]{\strut{}1.15}}}%
      \csname LTb\endcsname%
      \put(169,2045){\rotatebox{-270.00}{\makebox(0,0){\strut{}Speedup vs Grace ($\times$)}}}%
      \csname LTb\endcsname%
      \put(5398,224){\makebox(0,0){\strut{}NAS kernel (class D)}}%
      \csname LTb\endcsname%
      \put(1372,3533){\makebox(0,0)[l]{\strut{}Vera (88c)}}%
      \csname LTb\endcsname%
      \put(2907,3533){\makebox(0,0)[l]{\strut{}Vera (88c), SMT on}}%
      \csname LTb\endcsname%
      \put(5252,3533){\makebox(0,0)[l]{\strut{}EPYC 9R45 (96c)}}%
      \csname LTb\endcsname%
      \put(7442,3533){\makebox(0,0)[l]{\strut{}Xeon 6975P-C (96c)}}%
    }%
    \gplbacktext
    \put(0,0){\includegraphics[width={514.00bp},height={187.00bp}]{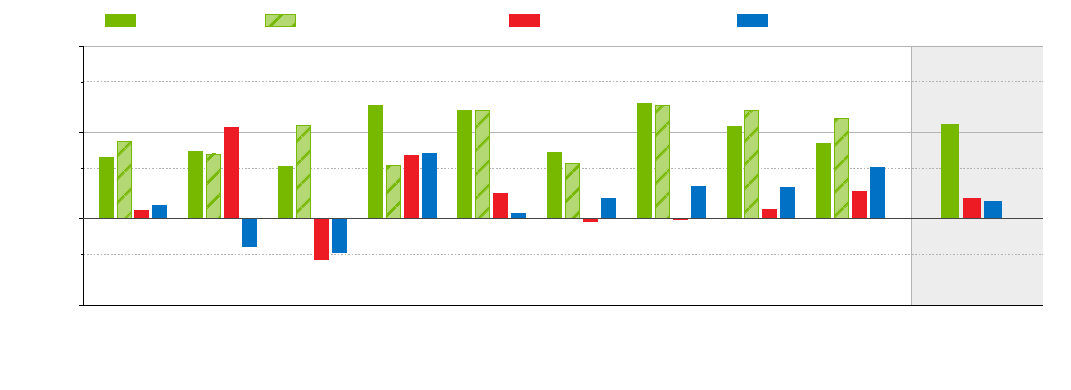}}%
    \gplfronttext
  \end{picture}%
\endgroup

  \caption{Speedup of each CPU across NAS Parallel Benchmark codes for class
  D input size when compared to an NVIDIA Grace full-chip baseline. Geometric mean over all kernels presents the faster of SMT off and SMT on.}
  \label{nas-speedup}
\end{figure*}

Figure~\ref{nas-speedup} presents the speedup in throughput, measured in \unit{\mega\op\per\second}, that each CPU achieves compared to a \num{72}-thread NVIDIA Grace baseline. We observe that Vera achieves the greatest speedup in all kernels except CG, a conjugate gradient kernel made up of unstructured matrix-vector multiplications. Surprisingly, AMD's 9R45 and Intel's 6975P-C performed poorly on EP in comparison to both NVIDIA Grace and Vera despite this kernel intending to estimate the floating point capabilities of the CPU, something which should demonstrate the larger vector lengths present on AMD and Intel's offerings. Vera's single-thread-per-core speedup compared to Grace here is expected, due to \vecunits{6}{128} SVE execute units on Vera compared to \vecunits{4}{128} SVE on Grace, leading to a \spd{1.5} increased theoretical FLOPs per cycle. Perf counter results reveal that Granite Rapids and Turin had low CPU utilization at \qty{53}{\percent} and \qty{61}{\percent} respectively, compared to \qty{99}{\percent} on Vera. We were unable to determine the cause for the low utilization as measured clock frequencies to be within their expected ranges.

We find that FT, MG, and IS codes yield the largest speedup for Vera. Characterization of each kernel by Brown and Jamieson finds that MG spends \qty{88}{\percent} of its runtime under high main-memory bandwidth utilization on Intel Skylake, and \qty{18}{\percent} for FT which also has frequent all-to-all collective inter core communication \cite{nas-characterisation}. We therefore attribute much of the \spd{2.52} speedup of MG compared to Grace to the \spd{2.42} all core achieved main-memory bandwidth. FT's \spd{2.49} speedup is due to \qty{22}{\percent} additional cores, \spd{2.42} greater main-memory bandwidth, and \qty{17}{\percent} reduced mean core-to-core latencies. IS makes frequent indirect memory accesses, so its \spd{2.39} speedup comes from Vera's lower cache latencies and its \qty{22}{\percent} more cores.

\begin{figure*}[t]
  \centering
  \input{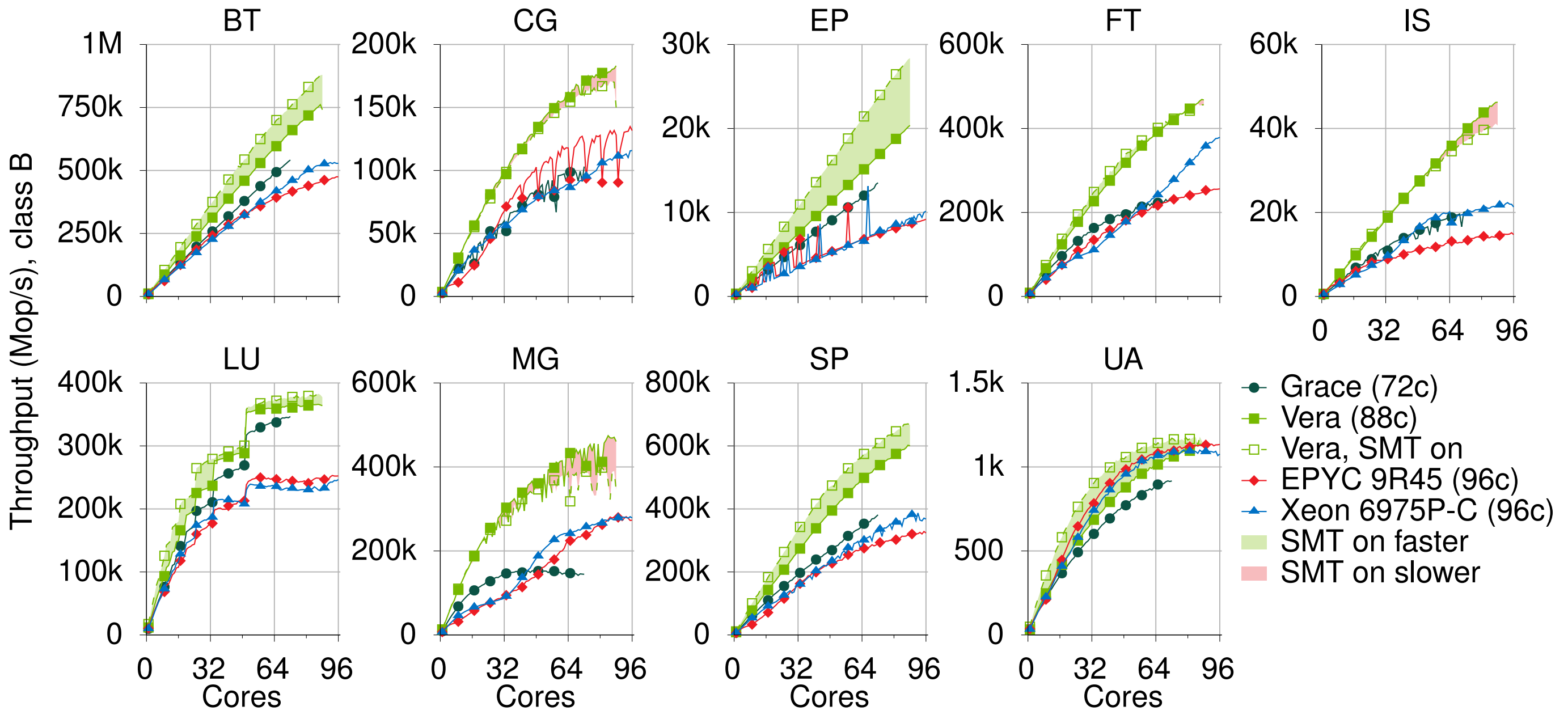}
  \caption{Scaling graph of \unit{\mega\op\per\second} achieved by each CPU on NAS Parallel Benchmarks with class B input size. The shaded band presents the difference between one versus two threads per core on Vera.}
  \label{nas-scaling}
\end{figure*}

Figure~\ref{nas-scaling} shows the strong scaling behavior for each NAS code compared to the other CPUs. Threads were incremented by sequential core IDs. We observe that Vera achieves the highest performance of CPUs tested at every core count in seven of the nine kernels, indicating both strong single-threaded performance and strong scaling that means that many of the kernels such as BT, SP, and CG, scale more linearly across the total core count compared to the x86 offerings which plateau more significantly as the number of cores increases. We note that while Turin achieved \spd{1.22} more \unit{\mega\op\per\second} than Vera on CG for class D, Vera outperforms Turin on class B. Perf results demonstrate that Vera has \spd{94.5} more dTLB misses per operation than Turin at \num{0.194} against \num{0.0021} for class D. This indicates that at class D the kernel is limited by address translation rather than bandwidth.

\begin{figure}[t]
  \centering
  \input{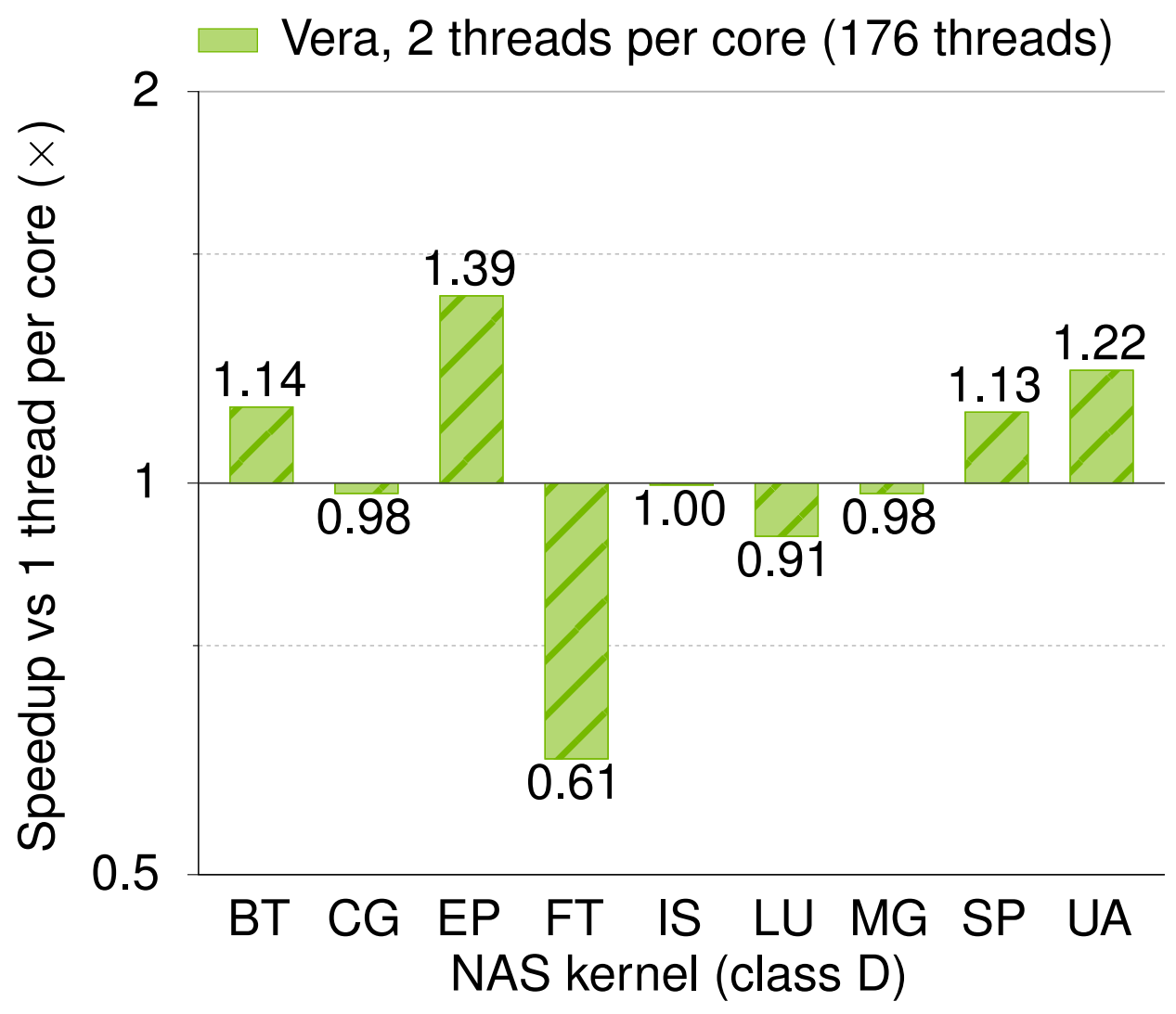}
  \caption{Speedup achieved across NAS Parallel Benchmark codes by NVIDIA
  Vera with two threads per core compared to a single thread per core.}
  \label{nas-smt}
\end{figure}

The effects of using all SMT threads are demonstrated in Figure~\ref{nas-smt} where we see a speedup in four of the NAS codes when compared to a single-thread-per-core baseline. EP leads to the greatest speedup at \spd{1.39}.  We infer that this is because the kernel has inter-thread independence allowing for strong scaling of threads, but each thread has a dependent chain of instructions leading to low ILP. An additional spatial multithread with half of the core's resources allows the exploitation of more ILP by saturating more of the core's resources as two independent instruction chains can run per core. Perf counter data supports our claim: SMT achieves \spd{1.43} increased IPC while maintaining similar clock frequency, cache miss rates, and branch misprediction rate.

SMT led to poor performance on FT with class D with \spd{0.61} the \unit{\mega\op\per\second} compared to running with a single thread per core. Interestingly, our class B results show roughly equal performance regardless of SMT use. This demonstrates two threads with independent data contending for the same cache, effectively halving the available cache per thread. We validate this with perf, finding that with SMT there are \spd{3.4} increased LLC-loads (L2 misses) while having \spd{5.5} LLC-misses, with the LLC miss rate increasing from \qty{20.4}{\percent} to \qty{32.8}{\percent}. The L1d miss-rate remains consistent at \qty{19}{\percent}. We infer that much of the data that was resident in L2 cache with a single thread per core misses both L2 and L3 cache in the two threads per-core case, increasing latency significantly. The default block factor of \num{32} leads to poor SMT performance for class D over \num{176} threads, while a factor of \num{8} leads to a \spd{1.13} improvement with two SMT threads per core.

\subsection{Miniapps}

Miniapps are reduced proxy applications based on common algorithms or derived from larger HPC codes. These proxy applications mimic the implementation of long-running portions of complex production codes and are used to deduce performance trends through careful experimentation. We utilize miniapps in this study to compare how each CPU's performance scales across multiple cores in a range of different compute patterns. We use four codes with different characteristics -- CloverLeaf \cite{cloverleaf}, TeaLeaf \cite{tealeaf}, miniBUDE \cite{minibude}, and Neutral \cite{neutral}. All miniapps were compiled and run with their OpenMP parallel implementation only.

CloverLeaf is a hydrodynamics miniapp that solves the compressible Euler equations in a 2D stencil grid. TeaLeaf solves the heat diffusion equations via a Conjugate Gradient linear solver on a 2D grid. Both are main memory-bandwidth bound on a single node; CloverLeaf's scaling is tightly coupled with the increase in aggregated memory bandwidth while TeaLeaf becomes communication-bound at scale due to the per-iteration dot-product reductions. Both are included in the SPEChpc 2021 suite of HPC benchmarks \cite{spechpc} and, more recently, SPEC CPU 2026 \cite{speccpu}.

MiniBUDE is a proxy application for BUDE, the Bristol University Docking Engine, a molecular docking code that predicts the structure formed between two molecules. This code is compute bound, with the L1d cache miss rate below \qty{0.1}{\percent} on each architecture. The performance of the OpenMP implementation is impacted by the poses per work-item (PPWI) parameter which aids compiler unrolling of the inner loops, leading to improved vectorization. We fix PPWI to \num{32} across all architectures as larger values would not decompose the fixed problem-size properly over large core counts, leading to work imbalance and thus CPU idle time.

Neutral is a Monte Carlo neutron transport miniapp. This code has data dependencies on the mesh that lead to random memory access patterns and poor data locality, which therefore leads the code to be memory latency bound. The impact of this memory latency varies depending on the input data, changing which particles interact and thus the memory access pattern.

\begin{figure}[t]
  \centering
  \begingroup
\fontfamily{phv}\selectfont
  \makeatletter
  \providecommand\color[2][]{%
    \GenericError{(gnuplot) \space\space\space\@spaces}{%
      Package color not loaded in conjunction with
      terminal option `colourtext'%
    }{See the gnuplot documentation for explanation.%
    }{Either use 'blacktext' in gnuplot or load the package
      color.sty in LaTeX.}%
    \renewcommand\color[2][]{}%
  }%
  \providecommand\includegraphics[2][]{%
    \GenericError{(gnuplot) \space\space\space\@spaces}{%
      Package graphicx or graphics not loaded%
    }{See the gnuplot documentation for explanation.%
    }{The gnuplot epslatex terminal needs graphicx.sty or graphics.sty.}%
    \renewcommand\includegraphics[2][]{}%
  }%
  \providecommand\rotatebox[2]{#2}%
  \@ifundefined{ifGPcolor}{%
    \newif\ifGPcolor
    \GPcolortrue
  }{}%
  \@ifundefined{ifGPblacktext}{%
    \newif\ifGPblacktext
    \GPblacktextfalse
  }{}%
  \let\gplgaddtomacro\g@addto@macro
  \gdef\gplbacktext{}%
  \gdef\gplfronttext{}%
  \makeatother
  \ifGPblacktext
    \def\colorrgb#1{}%
    \def\colorgray#1{}%
  \else
    \ifGPcolor
      \def\colorrgb#1{\color[rgb]{#1}}%
      \def\colorgray#1{\color[gray]{#1}}%
      \expandafter\def\csname LTw\endcsname{\color{white}}%
      \expandafter\def\csname LTb\endcsname{\color{black}}%
      \expandafter\def\csname LTa\endcsname{\color{black}}%
      \expandafter\def\csname LT0\endcsname{\color[rgb]{1,0,0}}%
      \expandafter\def\csname LT1\endcsname{\color[rgb]{0,1,0}}%
      \expandafter\def\csname LT2\endcsname{\color[rgb]{0,0,1}}%
      \expandafter\def\csname LT3\endcsname{\color[rgb]{1,0,1}}%
      \expandafter\def\csname LT4\endcsname{\color[rgb]{0,1,1}}%
      \expandafter\def\csname LT5\endcsname{\color[rgb]{1,1,0}}%
      \expandafter\def\csname LT6\endcsname{\color[rgb]{0,0,0}}%
      \expandafter\def\csname LT7\endcsname{\color[rgb]{1,0.3,0}}%
      \expandafter\def\csname LT8\endcsname{\color[rgb]{0.5,0.5,0.5}}%
    \else
      \def\colorrgb#1{\color{black}}%
      \def\colorgray#1{\color[gray]{#1}}%
      \expandafter\def\csname LTw\endcsname{\color{white}}%
      \expandafter\def\csname LTb\endcsname{\color{black}}%
      \expandafter\def\csname LTa\endcsname{\color{black}}%
      \expandafter\def\csname LT0\endcsname{\color{black}}%
      \expandafter\def\csname LT1\endcsname{\color{black}}%
      \expandafter\def\csname LT2\endcsname{\color{black}}%
      \expandafter\def\csname LT3\endcsname{\color{black}}%
      \expandafter\def\csname LT4\endcsname{\color{black}}%
      \expandafter\def\csname LT5\endcsname{\color{black}}%
      \expandafter\def\csname LT6\endcsname{\color{black}}%
      \expandafter\def\csname LT7\endcsname{\color{black}}%
      \expandafter\def\csname LT8\endcsname{\color{black}}%
    \fi
  \fi
    \setlength{\unitlength}{0.0500bp}%
    \ifx\gptboxheight\undefined%
      \newlength{\gptboxheight}%
      \newlength{\gptboxwidth}%
      \newsavebox{\gptboxtext}%
    \fi%
    \setlength{\fboxrule}{0.5pt}%
    \setlength{\fboxsep}{1pt}%
    \definecolor{tbcol}{rgb}{1,1,1}%
\begin{picture}(5020.00,4120.00)%
    \gplgaddtomacro\gplbacktext{%
      \csname LTb\endcsname%
      \put(644,800){\makebox(0,0)[r]{\strut{}0.5}}%
      \csname LTb\endcsname%
      \put(644,1467){\makebox(0,0)[r]{\strut{}1}}%
      \csname LTb\endcsname%
      \put(644,2135){\makebox(0,0)[r]{\strut{}2}}%
      \csname LTb\endcsname%
      \put(644,2802){\makebox(0,0)[r]{\strut{}4}}%
      \csname LTb\endcsname%
      \put(644,3470){\makebox(0,0)[r]{\strut{}8}}%
      \csname LTb\endcsname%
      \put(1284,651){\rotatebox{30.00}{\makebox(0,0)[r]{\strut{}CloverLeaf}}}%
      \csname LTb\endcsname%
      \put(2015,651){\rotatebox{30.00}{\makebox(0,0)[r]{\strut{}miniBUDE}}}%
      \csname LTb\endcsname%
      \put(2746,651){\rotatebox{30.00}{\makebox(0,0)[r]{\strut{}TeaLeaf}}}%
      \csname LTb\endcsname%
      \put(3477,651){\rotatebox{30.00}{\makebox(0,0)[r]{\strut{}Neutral}}}%
      \csname LTb\endcsname%
      \put(4427,651){\rotatebox{30.00}{\makebox(0,0)[r]{\strut{}}}}%
    }%
    \gplgaddtomacro\gplfronttext{%
      \csname LTb\endcsname%
      \put(4176,2372){\rotatebox{90.00}{\makebox(0,0)[l]{\strut{}2.41}}}%
      \csname LTb\endcsname%
      \put(4366,1950){\rotatebox{90.00}{\makebox(0,0)[l]{\strut{}1.56}}}%
      \csname LTb\endcsname%
      \put(4556,1734){\rotatebox{90.00}{\makebox(0,0)[l]{\strut{}1.24}}}%
      \csname LTb\endcsname%
      \put(169,2135){\rotatebox{-270.00}{\makebox(0,0){\strut{}Speedup vs Grace ($\times$)}}}%
      \csname LTb\endcsname%
      \put(4335,613){\makebox(0,0){\strut{}\raisebox{0pt}[\ht\strutbox][\dp\strutbox]{\begin{tabular}[t]{@{}c@{}}Geo. mean\\(best of\\SMT off/on)\end{tabular}}}}%
      \csname LTb\endcsname%
      \put(608,3928){\makebox(0,0)[l]{\strut{}Vera (88c)}}%
      \csname LTb\endcsname%
      \put(608,3685){\makebox(0,0)[l]{\strut{}Vera (88c), SMT on}}%
      \csname LTb\endcsname%
      \put(2948,3928){\makebox(0,0)[l]{\strut{}EPYC 9R45 (96c)}}%
      \csname LTb\endcsname%
      \put(2948,3685){\makebox(0,0)[l]{\strut{}Xeon 6975P-C (96c)}}%
    }%
    \gplbacktext
    \put(0,0){\includegraphics[width={251.00bp},height={206.00bp}]{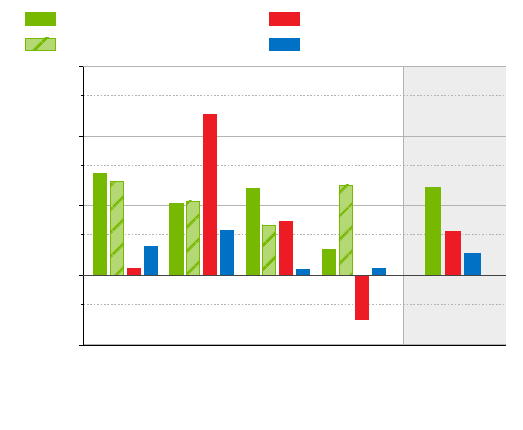}}%
    \gplfronttext
  \end{picture}%
\endgroup

  \caption{Speedup achieved by each CPU across HPC miniapp codes when compared to an NVIDIA Grace full-chip baseline. Geometric mean over all miniapps presents the faster of SMT off and SMT on.}
  \label{miniapp-speedup}
\end{figure}

Figure~\ref{miniapp-speedup} represents the speedup by each CPU compared to Grace on each miniapp. Vera achieves the highest performance on three of the four miniapps with a geometric mean of \spd{2.41} speedup. This is greatest in CloverLeaf which is \spd{2.76} faster than Grace. We can attribute most of this speedup to the \spd{2.42} increase in main memory bandwidth.

Turin achieves a \spd{2.38} speedup over Vera on miniBUDE. Vera issued \spd{3.91} more instructions per core compared to Turin, consistent with \qty{128}{\bit} vector instructions on Vera compared to \qty{512}{\bit} vector instructions on Turin. While Vera achieved \spd{1.81} higher IPC at \num{4.32}, Turin executed \spd{1.98} fewer cycles while also having more cores. Granite Rapids underperforms in miniBUDE which we uncover as a compiler performance issue due to \texttt{-march=native} leading to preferring a vector width of \qty{256}{\bit}, thus missing significant floating-point throughput in compute bound applications such as miniBUDE.

\begin{figure}[t]
  \centering
  \input{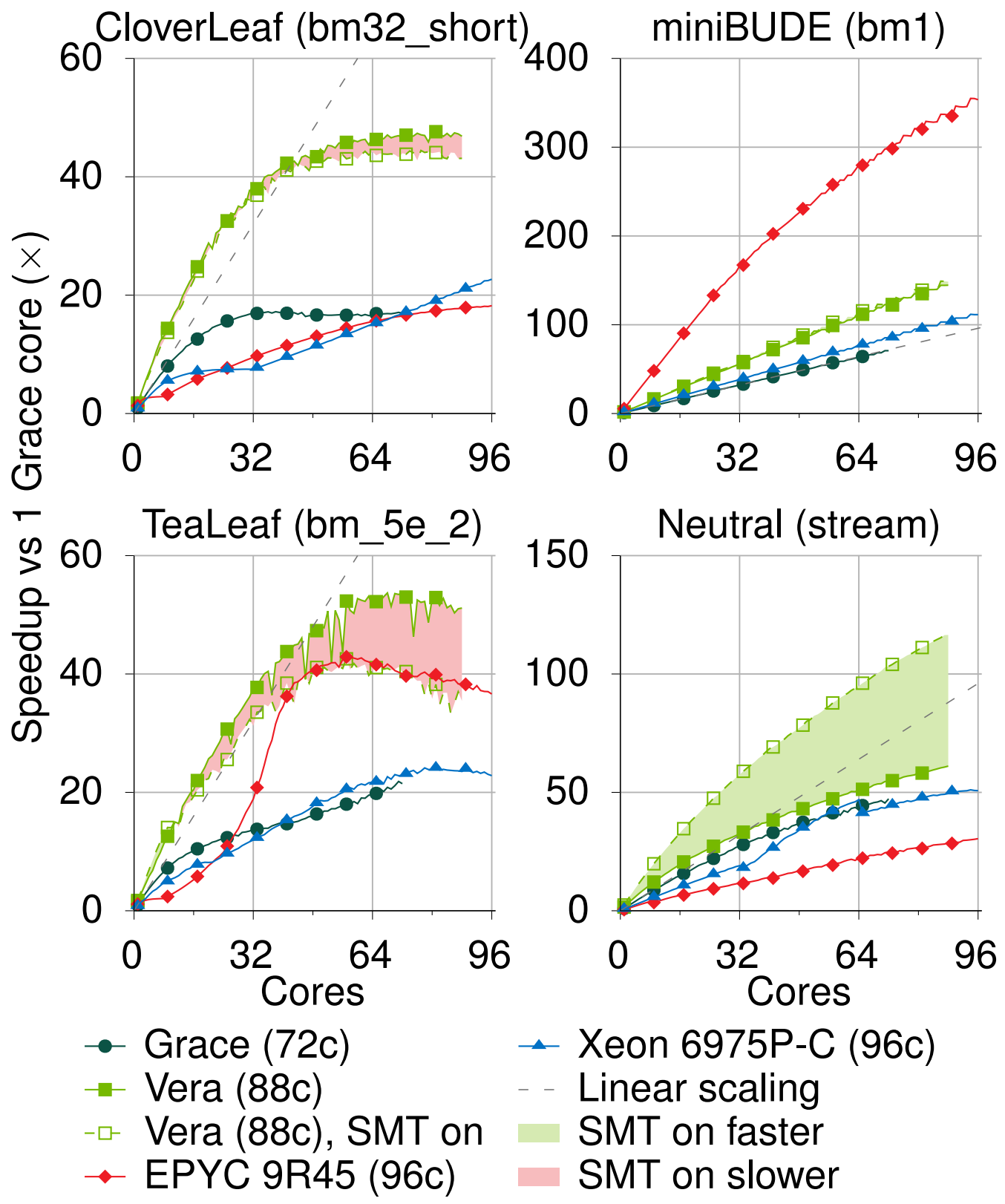}
  \caption{Scaling performance of each CPU across HPC miniapp codes when compared to an NVIDIA Grace single-core baseline. The shaded band presents the difference between one versus two threads per core on Vera.}
  \label{miniapp-scaling}
\end{figure}

Figure~\ref{miniapp-scaling} measures the strong scaling performance of the miniapps. Threads were again incremented by sequential core IDs. In CloverLeaf, TeaLeaf, and Neutral, Vera achieves the highest performance at all but one core count. In TeaLeaf and CloverLeaf, Vera reaches close to its highest performance more quickly than other architectures before plateauing as its memory bandwidth has been saturated. 

SMT causes a significant speedup to Neutral for Vera. Neutral has the lowest IPC on Vera of the four miniapps at one thread per core at \num{1.17}, reflecting its latency-bound profile. Introducing two threads per core increased the core's IPC to \num{2.31}, a \spd{1.97} increase, while the instruction count only increased by \spd{1.01}. SMT helps as the additional thread does not add significant memory pressure, with the miss rate for L1d and LLC remaining similar. This IPC gain is reflected in the \spd{1.91} speedup we see through running with two threads per core.

\subsection{Applications} \label{apps}

Application performance provides a ``real-world'' assessment of the types of work that are performed on high-end CPUs. Since Vera is not yet widely available, library dependencies are yet to be optimized for this architecture so results should not be considered a final reflection of performance. While the choice of performance libraries such as BLAS and FFT can have a noticeable impact on performance, we use one open-source software stack on all four CPUs such that a speedup reflects the hardware performance, not the maturity of the relevant vendor library. We therefore use \texttt{GCC 15.3.0}, \texttt{FFTW} \cite{fftw}, and \texttt{OpenBLAS} \cite{openblas} across all applications and architectures. Our Vera results therefore rely on optimizations targeting generic Armv8 or SVE targets and are likely to be conservative for BLAS and FFT-heavy codes such as CASTEP.

We select six applications due to their frequency of use in UK science; CASTEP \cite{castep}, CP2K \cite{cp2k}, GROMACS \cite{gromacs}, NAMD \cite{namd}, OpenFOAM \cite{openfoam} and WRF \cite{wrf}.  Each will show a particular performance characteristic of the code and are well studied and familiar to the communities.

Each application was built using Spack \cite{spack} version 1.1 and archspec was updated to automatically set the recommended compiler flags for Vera along with applying upstream fixes to Turin. Granite Rapids did not have support in archspec, and no upstream addition was available and was identified as Sapphire Rapids for compiler options. Spack was chosen for its flexibility and provides a consistency to the approach. Spack does not provide an out-of-the-box optimal build configuration and would require further work to tune the installation for each CPU outside the scope of this study.

\begin{figure*}[t]
  \centering
  \begingroup
\fontfamily{phv}\selectfont
  \makeatletter
  \providecommand\color[2][]{%
    \GenericError{(gnuplot) \space\space\space\@spaces}{%
      Package color not loaded in conjunction with
      terminal option `colourtext'%
    }{See the gnuplot documentation for explanation.%
    }{Either use 'blacktext' in gnuplot or load the package
      color.sty in LaTeX.}%
    \renewcommand\color[2][]{}%
  }%
  \providecommand\includegraphics[2][]{%
    \GenericError{(gnuplot) \space\space\space\@spaces}{%
      Package graphicx or graphics not loaded%
    }{See the gnuplot documentation for explanation.%
    }{The gnuplot epslatex terminal needs graphicx.sty or graphics.sty.}%
    \renewcommand\includegraphics[2][]{}%
  }%
  \providecommand\rotatebox[2]{#2}%
  \@ifundefined{ifGPcolor}{%
    \newif\ifGPcolor
    \GPcolortrue
  }{}%
  \@ifundefined{ifGPblacktext}{%
    \newif\ifGPblacktext
    \GPblacktextfalse
  }{}%
  \let\gplgaddtomacro\g@addto@macro
  \gdef\gplbacktext{}%
  \gdef\gplfronttext{}%
  \makeatother
  \ifGPblacktext
    \def\colorrgb#1{}%
    \def\colorgray#1{}%
  \else
    \ifGPcolor
      \def\colorrgb#1{\color[rgb]{#1}}%
      \def\colorgray#1{\color[gray]{#1}}%
      \expandafter\def\csname LTw\endcsname{\color{white}}%
      \expandafter\def\csname LTb\endcsname{\color{black}}%
      \expandafter\def\csname LTa\endcsname{\color{black}}%
      \expandafter\def\csname LT0\endcsname{\color[rgb]{1,0,0}}%
      \expandafter\def\csname LT1\endcsname{\color[rgb]{0,1,0}}%
      \expandafter\def\csname LT2\endcsname{\color[rgb]{0,0,1}}%
      \expandafter\def\csname LT3\endcsname{\color[rgb]{1,0,1}}%
      \expandafter\def\csname LT4\endcsname{\color[rgb]{0,1,1}}%
      \expandafter\def\csname LT5\endcsname{\color[rgb]{1,1,0}}%
      \expandafter\def\csname LT6\endcsname{\color[rgb]{0,0,0}}%
      \expandafter\def\csname LT7\endcsname{\color[rgb]{1,0.3,0}}%
      \expandafter\def\csname LT8\endcsname{\color[rgb]{0.5,0.5,0.5}}%
    \else
      \def\colorrgb#1{\color{black}}%
      \def\colorgray#1{\color[gray]{#1}}%
      \expandafter\def\csname LTw\endcsname{\color{white}}%
      \expandafter\def\csname LTb\endcsname{\color{black}}%
      \expandafter\def\csname LTa\endcsname{\color{black}}%
      \expandafter\def\csname LT0\endcsname{\color{black}}%
      \expandafter\def\csname LT1\endcsname{\color{black}}%
      \expandafter\def\csname LT2\endcsname{\color{black}}%
      \expandafter\def\csname LT3\endcsname{\color{black}}%
      \expandafter\def\csname LT4\endcsname{\color{black}}%
      \expandafter\def\csname LT5\endcsname{\color{black}}%
      \expandafter\def\csname LT6\endcsname{\color{black}}%
      \expandafter\def\csname LT7\endcsname{\color{black}}%
      \expandafter\def\csname LT8\endcsname{\color{black}}%
    \fi
  \fi
    \setlength{\unitlength}{0.0500bp}%
    \ifx\gptboxheight\undefined%
      \newlength{\gptboxheight}%
      \newlength{\gptboxwidth}%
      \newsavebox{\gptboxtext}%
    \fi%
    \setlength{\fboxrule}{0.5pt}%
    \setlength{\fboxsep}{1pt}%
    \definecolor{tbcol}{rgb}{1,1,1}%
\begin{picture}(10280.00,3960.00)%
    \gplgaddtomacro\gplbacktext{%
      \csname LTb\endcsname%
      \put(644,716){\makebox(0,0)[r]{\strut{}0.5}}%
      \csname LTb\endcsname%
      \put(644,1647){\makebox(0,0)[r]{\strut{}1}}%
      \csname LTb\endcsname%
      \put(644,2577){\makebox(0,0)[r]{\strut{}2}}%
      \csname LTb\endcsname%
      \put(644,3508){\makebox(0,0)[r]{\strut{}4}}%
      \csname LTb\endcsname%
      \put(1393,530){\makebox(0,0){\strut{}\raisebox{0pt}[\ht\strutbox][\dp\strutbox]{\begin{tabular}[t]{@{}c@{}}CASTEP\\(al3x3)\end{tabular}}}}%
      \csname LTb\endcsname%
      \put(2597,530){\makebox(0,0){\strut{}\raisebox{0pt}[\ht\strutbox][\dp\strutbox]{\begin{tabular}[t]{@{}c@{}}CP2K\\(fayalite)\end{tabular}}}}%
      \csname LTb\endcsname%
      \put(3802,530){\makebox(0,0){\strut{}\raisebox{0pt}[\ht\strutbox][\dp\strutbox]{\begin{tabular}[t]{@{}c@{}}GROMACS\\(TestCaseB)\end{tabular}}}}%
      \csname LTb\endcsname%
      \put(5006,530){\makebox(0,0){\strut{}\raisebox{0pt}[\ht\strutbox][\dp\strutbox]{\begin{tabular}[t]{@{}c@{}}NAMD\\(stmv)\end{tabular}}}}%
      \csname LTb\endcsname%
      \put(6211,530){\makebox(0,0){\strut{}\raisebox{0pt}[\ht\strutbox][\dp\strutbox]{\begin{tabular}[t]{@{}c@{}}OpenFOAM$^{*}$\\(hpc\_motorbike)\end{tabular}}\kern8pt}}%
      \csname LTb\endcsname%
      \put(7415,530){\makebox(0,0){\strut{}\raisebox{0pt}[\ht\strutbox][\dp\strutbox]{\begin{tabular}[t]{@{}c@{}}WRF\\(conus12k)\end{tabular}}}}%
      \csname LTb\endcsname%
      \put(9041,530){\makebox(0,0){\strut{}\raisebox{0pt}[\ht\strutbox][\dp\strutbox]{\begin{tabular}[t]{@{}c@{}}Geo. mean\\(best of SMT off/on)\end{tabular}}}}%
    }%
    \gplgaddtomacro\gplfronttext{%
      \csname LTb\endcsname%
      \put(8807,2557){\rotatebox{90.00}{\makebox(0,0)[l]{\strut{}1.89}}}%
      \csname LTb\endcsname%
      \put(9072,2142){\rotatebox{90.00}{\makebox(0,0)[l]{\strut{}1.39}}}%
      \csname LTb\endcsname%
      \put(9337,2431){\rotatebox{90.00}{\makebox(0,0)[l]{\strut{}1.72}}}%
      \csname LTb\endcsname%
      \put(169,2113){\rotatebox{-270.00}{\makebox(0,0){\strut{}Speedup vs Grace ($\times$)}}}%
      \csname LTb\endcsname%
      \put(687,3753){\makebox(0,0)[l]{\strut{}Vera (88c)}}%
      \csname LTb\endcsname%
      \put(2223,3753){\makebox(0,0)[l]{\strut{}Vera, SMT on}}%
      \csname LTb\endcsname%
      \put(4057,3753){\makebox(0,0)[l]{\strut{}EPYC 9R45 (96c)}}%
      \csname LTb\endcsname%
      \put(6247,3753){\makebox(0,0)[l]{\strut{}Xeon 6975P-C (96c)}}%
      \csname LTb\endcsname%
      \put(8670,3753){\makebox(0,0)[l]{\strut{}Xeon, SMT on}}%
    }%
    \gplbacktext
    \put(0,0){\includegraphics[width={514.00bp},height={198.00bp}]{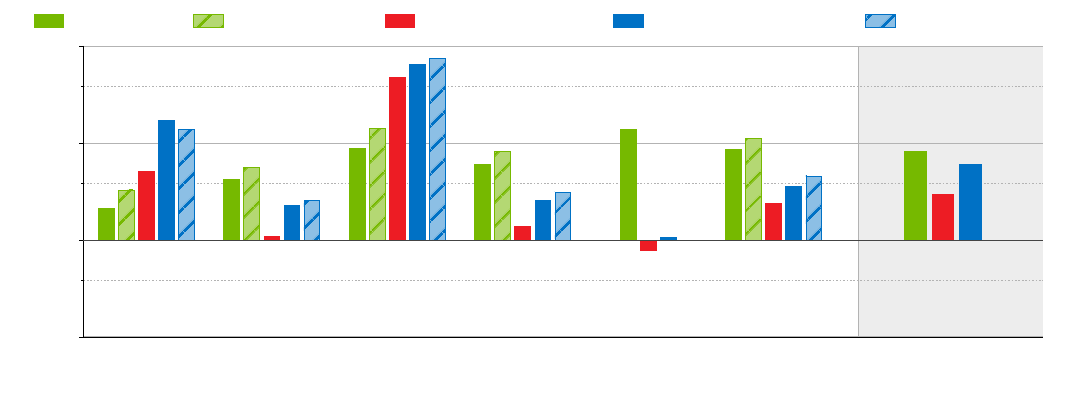}}%
    \gplfronttext
  \end{picture}%
\endgroup

  \caption{Speedup achieved by each CPU across scientific applications when
  compared to an NVIDIA Grace full-chip baseline. Geometric mean over all applications presents the faster of SMT off and SMT on. $^{*}$Native threading is unsupported in OpenFOAM, so only one rank per core is presented.}
  \label{apps-speedup}
\end{figure*}

Each benchmark case fully populated the CPU with the most appropriate method for a single socket using Slurm (e.g. to stop distributing tasks across sockets). For most codes this resulted in MPI tasks across all CPU cores and when testing multithreading, OpenMP was used to populate any extra hardware threads (SMT) on a CPU core.  The summary of results for each CPU and application is shown in Figure~\ref{apps-speedup}.

CASTEP calculates material properties using plane-waves density-functional theory. It is closed-source, but free for academic use across the world. The al3x3 case is small enough to fit into main-memory across each system. The results suggest the choice of FFT and BLAS libraries may be driving the lower than expected performance for Vera, which achieves \spd{1.25} over Grace with one thread per core and \spd{1.43} with two, and for Turin at \spd{1.63}, against \spd{2.34} for Granite Rapids. We used perf to find that \texttt{zgemm} kernels were dominant and their implementation would drive the performance characteristic. Intel, AMD and NVIDIA all provide optimized BLAS libraries that should benefit this code.

CP2K is a package for atomistic simulation. It supports ab initio electronic structure and classical force-field molecular dynamics. The Fayalite-FIST case runs classical molecular dynamics of \qty{28000} atoms with the FIST force-field module for 1000 steps. This input writes coordinates and velocities every step and a restart every \qty{20} steps on a single rank, leading to \qty{62}{\percent} of the runtime being dominated by IO. Wall time therefore measures file output more than the CPU architecture. We instead report the force-evaluation time (\texttt{fist\_calc\_energy\_force} maximized over all ranks, as reported in the code's output file) which is the primary kernel for this input case. Vera evaluates forces \spd{1.54} faster than Grace with one thread per core or \spd{1.68} with two. Turin and Granite Rapids report a \spd{1.03} and \spd{1.28} speedup respectively despite their wider vector units, which suggests this routine is not limited by vector floating-point throughput.

GROMACS is an extensively used molecular dynamics package that solves Newton’s equations of motion. Systems of interest, such as proteins, can contain millions of particles. At low node counts it is considered to be FLOP/s-bound. We used perf to identify the SIMD kernels provided by GROMACS as the dominant factor. The performance of GROMACS is seen to benefit the Intel and AMD CPUs due to the well optimized FLOP/s performance whilst other CPUs compatible with the Arm ISA have known issues with maximizing FLOP/s. We found that NEON provided improved performance over SVE instructions for Arm and is the recommendation in GROMACS documentation.

NAMD is a molecular dynamics simulation program designed to scale up to millions of atoms. The multicore version of NAMD was built to simplify the runtime configuration on a single socket. Similar to CP2K, many factors define performance. Vera's \spd{2.42} higher memory bandwidth and overall microarchitecture improvements drive its \spd{1.88} speedup compared to Grace.

OpenFOAM is a modular C++ framework that aims to simplify writing custom computational fluid dynamics (CFD) solvers. It is considered memory-bound on a single node.  As there is no shared-memory threading option in the main distribution, we did not experiment with oversubscribing MPI processes and thus left SMT investigation for future work. The code takes advantage of the available memory bandwidth in Vera and achieves a speedup of \spd{2.20}, similar to the increase in bandwidth. Intel and AMD have closer memory bandwidth to Grace and therefore have reduced gains in performance.

WRF is a mesoscale numerical weather prediction system. We report the mean wall time per model step, excluding the first step which includes initialization. We use the CONUS \qty{12}{\kilo\meter} case for \qty{600} time steps. Of these, \qty{57} also compute radiation, where these steps take approximately \spd{9} the time of other steps, accounting for roughly half of the total step time. Vera is \spd{2.05} faster than Grace on non-radiation steps and \spd{1.78} on the radiation steps, leading to \spd{1.91} overall and \spd{2.07} with two threads per core. WRF is partially memory-bound on a single node, while radiation steps depend more on floating point throughput. Granite Rapids achieves a \spd{1.77} speedup on non-radiation steps with only \spd{1.23} Grace's STREAM bandwidth, suggesting that per-core compute throughput also contributes to performance.

Overall, Vera achieves a significant speedup over our selection of applications at \spd{1.89} of Grace when considering the best time out of SMT on/off. While this result is higher than the \spd{1.72} speedup Granite Rapids achieves, Granite Rapids' results are skewed upwards due to its very strong performance on CASTEP and GROMACS, while Vera presents large performance improvements when compared to Grace throughout. Many FLOP/s dominated codes, including GROMACS and CASTEP, have GPU implementations that provide significantly higher floating-point throughput. Further work on utilizing optimized vendor libraries with each application would also likely improve performance for each application.

\section{Conclusion} \label{conclusion}

This work has provided a holistic analysis of NVIDIA Vera through the use of microbenchmarks, miniapps, and scientific applications. We base our claims in comparison to NVIDIA Grace, while providing an insight into x86 offerings through the inclusion of AMD Turin and Intel Granite Rapids. Overall, we find that Vera is the strongest CPU out of our selection of CPUs. Vera provides a significant generational performance improvement compared to Grace; when comparing geometric means, we observe a \spd{2.13} speedup in NAS Parallel Benchmarks, \spd{2.41} speedup across our miniapps, and \spd{1.89} speedup in our scientific applications. While Turin and Granite Rapids have competing results in select applications, such as a \spd{1.65} for Granite Rapids against Vera for GROMACS, and \spd{2.38} for Turin against Vera in miniBUDE, Vera remains the highest performing option for the majority of codes tested.

Our methodology helps to uncover why we observe such speedups from Vera. Vera's \qty{1088.2}{\giga\byte\per\second} sustained memory bandwidth, \spd{2.42} what was achieved on Grace, contributes significantly to the performance uplift of memory bound applications.

Spatial Multithreading benefits the majority of the scientific workloads tested yielding a \spd{1.11} geometric mean speedup. We find four of nine NAS codes have a performance improvement, all of which are between \spd{1.13} and \spd{1.39}. In the most extreme case of a highly latency-bound miniapp, Neutral, we find a \spd{1.91} speedup through using SMT. While Intel's Hyper-Threading model provides improvement over many of the real-world applications, the magnitude of this improvement is often weaker, achieving a \spd{1.05} speedup in the scientific codes tested where it provided a performance gain.

Vera's lowest average core-to-core latencies across its monolithic compute die and low cache latencies, aided by prefetching select irregular data layouts, further enable Vera to scale performance across its full core count effectively on most workloads, particularly communication-heavy or latency bound codes.

Vera primarily struggles in two cases: First, its modest \vecunits{6}{128} SVE vector units lead to lower floating point performance per cycle while having fewer cores which is visible in compute-bound applications such as miniBUDE and GROMACS. Second, while adoption of the Arm ISA has grown in recent years, performance tuning for long-standing applications is still catching up. For example, GROMACS performance suffers compared to x86 and likely requires further algorithmic changes or more architectural optimizations to close the gap for some Arm-compatible platforms. We expect that as Arm becomes more mainstream in HPC, this will broaden optimization efforts across the software ecosystem, further enhancing performance on platforms such as Vera.

Overall, we believe Vera to be an impressive data-center CPU that is not limited to iterative improvements over its predecessor Grace, but instead makes significant leaps in multiple areas of the core and chip design that provide strong performance across the majority of tested workloads when compared to Grace, Turin, and Granite Rapids. Future work should explore how Vera performs against next-generation AMD and Intel offerings, Venice and Diamond Rapids respectively. Alternatively, a deeper dive into Vera's core architecture including specifics of the Spatial Multithreading implementation would help understand further where scientific codes can exploit this microarchitecture fully.

\section*{AI Disclaimer}

Large Language Models, specifically Claude Opus 5, Opus 5.5, and Sonnet 5, were used to aid the creation of code to generate run scripts, build scripts, scripts to help query and interpret results, and plotting scripts. These are included within the Artifact Description (AD). Within the paper text, they were only used during the review stage to help strengthen and condense text, spot inconsistencies in writing, as well as for LaTeX formatting. These tools were not used for conceptual ideas, result analysis, or citations. The authors edited all text output carefully and take full accountability for the text provided.

\section*{Acknowledgements}

The authors would like to thank NVIDIA for providing pre-release access to engineering samples of Vera, as well as for accommodating a visit to NVIDIA HQ to learn more about the Olympus core architecture. In particular, we highlight the support from Giridhar Chukkapalli, Filippo Spiga, Dan Ernst, and Ananya Chakraborty.

The authors acknowledge the use of resources provided by the Isambard 3 Tier-2 HPC Facility. Isambard 3 is hosted by the University of Bristol and operated by the GW4 Alliance (https://gw4.ac.uk) and is funded by UK Research and Innovation; and the Engineering and Physical Sciences Research Council [EP/X039137/1].

\clearpage
   \includepdf[pages=-]{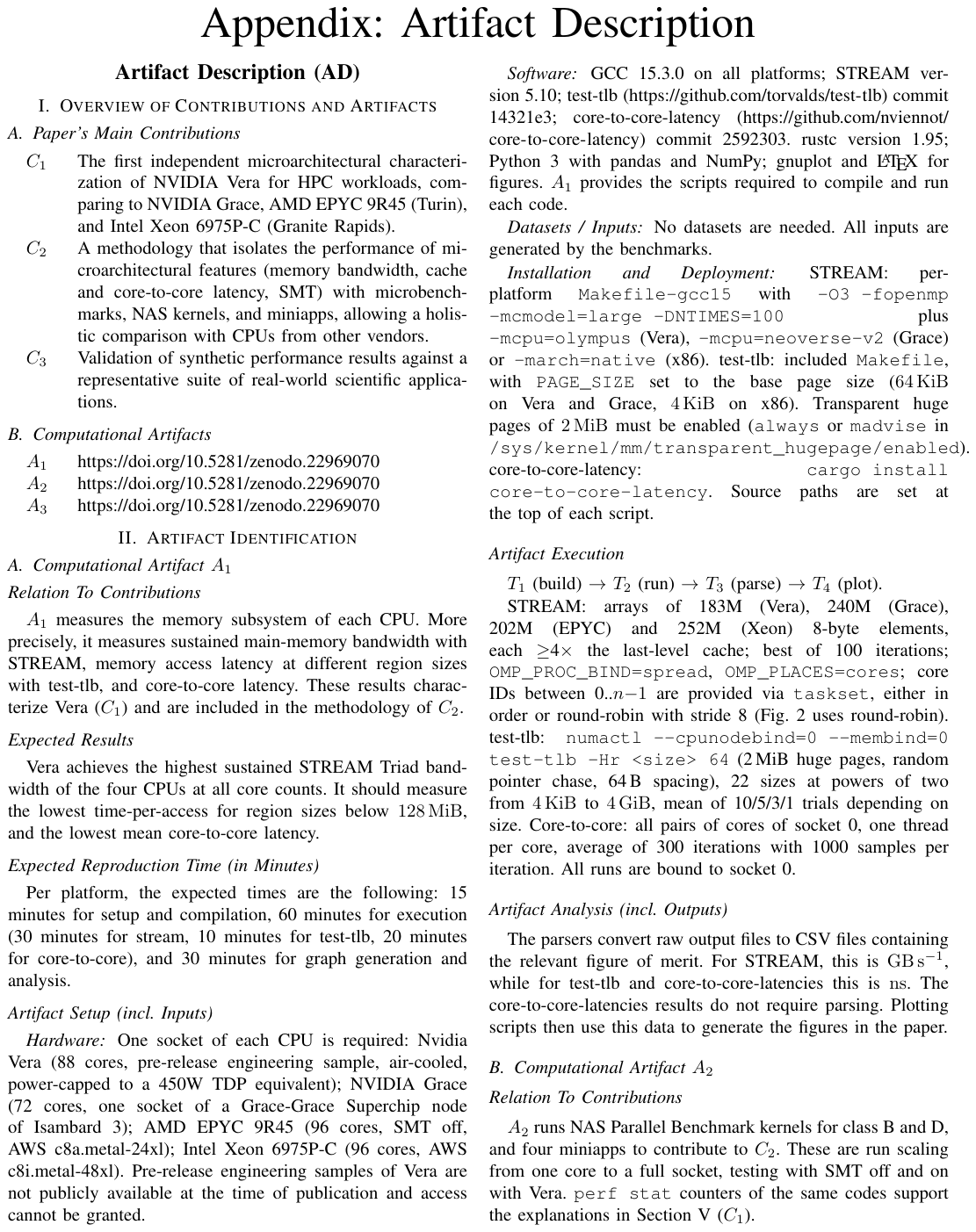}
   
\end{document}